\documentclass{aa}  
\bibpunct{(}{)}{;}{a}{}{,} % to follow the A&A style
\usepackage{natbib}
\usepackage{graphicx}
\usepackage{txfonts}
\begin{document} 

  \title{Terrestrial-type planet formation:} 
   \subtitle{Comparing different types of initial conditions}

   \author{M. P. Ronco\thanks{mpronco@fcaglp.unlp.edu.ar}, G. C. de El\'{\i}a \& O. M. Guilera}

\offprints{M. P. Ronco}

   \institute{Facultad de Ciencias Astron\'omicas y Geof\'\i sicas, Universidad
Nacional de La Plata and Instituto de Astrof\'{\i}sica de La Plata, CCT La Plata-CONICET-UNLP,
   Paseo del Bosque S/N (1900), La Plata, Argentina}

   \date{Received / Accepted }
% \abstract{}{}{}{}{} 
% 5 {} token are mandatory
%ABSTRACT - VERSION 1
  \abstract
  % leave it empty if necessary  
      { In order to study the terrestrial-type planet formation during the post oligarchic growth, the initial distributions
        of planetary embryos and planetesimals used in N-body simulations play an important role. In general, most of these studies 
        typically use ad hoc initial distributions based primarily on theoretical and numerical studies.} 
      {We analyze the formation of planetary systems without gas giants around solar-type stars focusing
        on the sensitivity of the results to the particular initial distributions used for planetesimals
        and planetary embryos at the end of the gas phase of the protoplanetary disk. The formation process
        of terrestrial planets in the habitable zone (HZ) and the study of their final water contents are also
        topics of special interest in this work.}
      {We developed two different sets of N-body simulations starting with the same protoplanetary disk. 
        The first set assumes typical ad hoc initial distributions for embryos and planetesimals and the second set obtains 
        these initial distributions from the results of a semi-analytical model which simulates the evolution of the 
        protoplanetary disk during the gaseous phase.}
      {Both sets of simulations form planets within the HZ. Using ad hoc initial conditions the masses of the planets that 
        remain in the HZ range from $0.66M_\oplus$ to $2.27M_\oplus$. Using more realistic initial conditions obtained from a 
        semi-analytical model, we found that the masses of the planets range from $1.18M_\oplus$ to $2.21M_\oplus$. Both sets of simulations form 
        planets in the HZ with water contents ranging between 4.5\% and 39.48\% by mass. Those planets that have the highest water 
        contents with respect to those with the lowest water contents, present differences regarding the sources of water supply.}
      {Comparing both sets of simulations we suggest that the number of planets that remain in the HZ is not sensitive to
        the particular initial distribution of embryos and planetesimals and thus, the results are globally
        similar between them. However, the main differences observed between both sets are associated to the 
        accretion history of the planets in the HZ. These discrepancies have a direct impact in the accretion of 
        water-rich material and in the physical characteristics of the resulting planets.}

                             
   \maketitle
%________________________________________________________________

\section{Introduction}
The formation process of terrestrial-type planets goes through different stages.
In the classical core accretion model \citep{LissawerStevenson2007}, the first step in the planet formation process is the 
sedimentation to the mid-plane of the dust of the disk and the formation of planetesimals. The mechanisms which lead to the 
formation of planetesimals are still under debate. However, in the last few years, three main scenarios for the formation of 
planetesimals have emerged: formation via coagulation-fragmentation cycles and mass transfer from small to large aggregates, 
growth of fluffy particles and subsequent compaction by self-gravity, and concentration of small particles (often called 
pebbles) in the turbulent gas, and gravitational fragmentation of over dense filaments (see \citealp{Johansen2014} for a detailed
 discussion and references therein). After the formation of the first population of planetesimals, these bodies grow 
quickly by mutual accretion until some of them (often called embryos) start to separate from the population of planetesimals. 
This process is known as planetesimal runaway growth. When embryos begin to gravitationally dominate the dynamic of the population 
of planetesimals, the runaway growth regime switches to the oligarchic growth regime. In this regime, the gravitational excitation 
of the embryos over the planetesimals limits their growth, and embryos are the only population that continue growing by accretion of 
planetesimals \citep{Kokubo1998}. 

The interaction of the embryos with the gaseous disk could lead to the 
migration of these bodies along the disk, a phenomenon known as type I migration. In idealized isothermal disks type I migration 
is always inward and at high migration rates \citep{Tanaka2002}. However, if more realistic disks are considered 
\citep{Paardekooper2010, Paardekooper2011} type I migration could be outward. This outward migration can also occur when the planet achieves 
a significant mass ($\sim 10M_{\oplus}$) and corrotation torques become not negligible \citep{Masset2006}. But even
in isothermal disks, type I migration could be outward if full magnetohydrodynamic disk (MHD) turbulence is considered \citep{Guilet2013}. More recently, 
\citet{BenitezLlambay2015} found that if the released energy by the planet due to accretion of solid material is taken into 
account this phenomena generates a \emph{heating torque} which could significantly slow down, cancel, or even reverse, inward type 
I migration for planets with masses $\lesssim 5M_{\oplus}$. Embryos continue growing by accretion of planetesimals and other embryos when 
the distance between them becomes less than 3-4 mutual Hill radii \citep{AgnorAsphaug2004} due to 
the fact that the damping generated by the gas remains the embryos in nearly circular and coplanar orbits. 

It is worth mentioning that in the last years a new alternative model was proposed for the formation of giant planet cores and 
terrestrial planets. This model proposes that the core of giant planets and terrestrial planets could be formed as seeds that accrete 
small particles, often called pebbles. \citet{Ormel2010b}, \citet{Lambrechts2012}, and \citet{Guillot2014} 
showed that pebbles, which are strong coupled to the gas, could be accreted very efficiently to 
form planetary bodies. 

More recently, \citet{Johansen2015} showed that other structures of 
our Solar System, like the asteroid belt and the transneptunian belt, could also be explained by the pebble accretion mechanism.     

Once the gas has dissipated from the system, and maybe some giant planet have formed, the terrestrial planet region contains a great 
number of planetary embryos which still evolve embedded in a remnant swarm of planetesimals. Embryos continue growing by accreting 
planetesimals and by stochastic collisions with other embryos in a regime known as pos-oligarchic growth. Finally, the 
terrestrial-type planets form and settle onto stable orbits on time scales of the order of 10$^{8}$ years.

In general terms, the numerical simulations that analyze the late stage planetary
accretion after the gas disk dissipation require physical and orbital
initial conditions for planetesimals and planetary embryos, as well as for the gas giants
of the system. These initial conditions are tipically selected ad hoc by many works that investigate the process of planetary formation
\citep{OBrien2006,Raymond2009,Morishima2010, Walsh2011}.

The study of planetary systems without gas giants results to be very interesting. In fact,
several observational studies \citep{Cumming2008,Mayor2011} and theoretical works
\citep{Mordasini2009,Miguel2011} have suggested that the planetary systems
consisting only of rocky planets would seem to be the most common in the Universe. Recently,
\citet{RoncodeElia2014} have analyzed the terrestrial-type planet formation and water delivery
in systems without gas giants around Sun-like stars. In particular, the authors focused on low-mass
protoplanetary disks for a wide range of surface density profiles. In that work, \citet{RoncodeElia2014}
assumed \emph{ad hoc} initial conditions for the distributions of planetesimals and planetary embryos after the gas dissipation.

Here, we present results of N-body simulations aimed at analyzing the process of planetary formation
around Sun-like stars in absence of gas giants. In particular, the present study focuses on the sensitivity
of the results to the particular initial distribution adopted for planetesimals and planetary embryos
after the gas dissipation. To do this, we decided to carry out two different sets of numerical simulations
starting with the same initial protoplanetary disk and modifying the particular initial distribution
for planetesimals and planetary embryos. The first set of numerical simulations assumes ad hoc
initial conditions for planetesimals and embryos like in \citet{RoncodeElia2014} while the second set of 
numerical simulations obtains initial conditions from a semi-analytical model, which is
able to analyze the evolution of a system of planetesimals and embryos during the gaseous phase.
A comparative analysis between these two sets of numerical simulations will allow us to enrich our knowledge
concerning the process of planetary formation. Moreover, we consider that the present work will allow us
to clarify our understanding about a correct selection of initial conditions for numerical simulations
of formation and evolution of planetary systems.

This paper is therefore structured as follows. In Sect. 2, we present the main properties of
the protoplanetary disk used in our simulations. Then, we outline our choice of initial conditions in Sect. 3 and
discuss the main characteristics of the N-body code in Sect. 4. In Sect. 5, we
present the results of all our simulations. We discuss such results within the framework
of current knowledge of planetary systems in Sect. 6 and present our conclusions in Sect. 7.
%__________________________________________________________________
 \section{Protoplanetary Disk: Properties}

The properties of the protoplanetary disk considered in this work are the same presented in \citet{RoncodeElia2014}.
The gas surface density profile that represents the structure of the protoplanetary disk is given by
\begin{eqnarray}
\Sigma_{\text{g}}(r) = \Sigma^{0}_{\text{g}}\left(\dfrac{r}{r_{\text{c}}}\right)^{-\gamma}e^{-(\frac{r}{r_{\text{c}}})^{2-\gamma}},
\label{eq:gas}
\end{eqnarray}
where $\Sigma_{\text{g}}^{0}$ is a normalization constant, $r_{\text{c}}$ a characteristic radius and $\gamma$ the
exponent that represents the surface density gradient. $\Sigma_{\text{g}}^{0}$ can be found by integrating Eq.
\ref{eq:gas} over the total area of the disk and is a function of $\gamma$, $r_{\text{c}}$ and the mass of the disk $M_{\text{d}}$.

Analogously, the solid surface density profile $\Sigma_{\text{s}}(r)$ is represented by
\begin{eqnarray}
\Sigma_{\text{s}}(r) = \Sigma^{0}_{\text{s}}\eta_{\text{ice}}\left(\dfrac{r}{r_{\text{c}}}\right)^{-\gamma}e^{-(\frac{r}{r_{\text{c}}})^{2-\gamma}},
\label{eq:solidos}
\end{eqnarray}
where $\eta_{\text{ice}}$ represents an increase of a factor of 4 in the amount of solid material due to the condensation
of water beyond the snow line, which is located at 2.7~au \citep{Hayashi1981}. Although other authors found a much lower 
value for this increase of solids \citep{Lodders2003} we used the same factor as \citet{RoncodeElia2014} in order to be consistent
with the comparisons.

The relation between both profiles is given by
\begin{eqnarray}
\left(\dfrac{\Sigma^{0}_{\text{s}}}{\Sigma^{0}_{\text{g}}}\right)_{\star} = \left(\dfrac{\Sigma^{0}_{\text{s}}}{\Sigma^{0}_{\text{g}}}\right)_{\odot}10^{[\text{Fe}/\text{H}]} = z_{0}10^{[\text{Fe}/\text{H}]},
\end{eqnarray}
where $z_0$ is the primordial abundance of heavy elements in the Sun and $[\text{Fe}/\text{H}]$ the metallicity.

The central star is assumed to have the same mass as our Sun and is a solar metallicity star, so that $[\text{Fe}/\text{H}]=0$.
Then, $\Sigma^{0}_{\text{s}} = z_{0}\Sigma^{0}_{\text{g}}$
where $z_0 = 0.0149$ \citep{Lodders2003}. We consider a low-mass disk with $M_{\text{d}} = 0.03M_\odot$ in order to guarantee
that the formation of giant planets is not performed \citep{Miguel2011}.
For the characteristic radius $r_{\text{c}}$ we adopt a value of 50 au which is in agreement with the different disk's observations 
studied by \citet{Andrews2010}, and for the exponent $\gamma$ we adopt a value of 1.5 according to other planetary formation works 
\citep{Raymond2005,Miguel2011}. The gas surface density profile of this disk is a factor of $1-1.7$ higher than that proposed by the Minimun Mass Solar Nebula (MMSN) between 0 and 15 au. It is worth noting that the simulations
developed by \citet{RoncodeElia2014} using density profiles with $\gamma = 1.5$ show distinctive results since true water worlds
are formed in the habitable zone\footnote{ The definition of the habitable zone is postponed to Sec. 5 but we anticipate that
we consider a conservative and an optimistic habitable zone as in \citet{Kopparapuerrata2013,Kopparapu2013}.} (HZ) of the system .

Finally, we assumed that the protoplanetary disk presents a radial compositional gradient. Thus, the water content by mass $W(r)$ assumed
as a function of the radial distance $r$ is described as: embryos and planetesimals contain $0.001\%$ water inside
2~au, contain $0.1\%$ water between 2~au and 2.5~au, contain $5\%$ water between
2.5~au and 2.7~au, and $50\%$ water past 2.7~au. This water distribution is the same used by \citet{RoncodeElia2014} and
is based on that proposed by \citet{Raymond2004}. Such a distribution is assigned to each body in the N-body simulations,
based on its starting location. Our model does not consider water loss during impacts, thus, the final water contents of the
resulting planets will represent upper limits.

\section{Initial conditions}

As mentioned in Sec. 1, many works that investigate the process of planetary formation 
consider ad hoc initial conditions, particularly dividing the solid mass of the disk equally between the planetesimal and
embryo components. Moreover, some works \citep{Chambers2001,OBrien2006} assume that
the individual mass of the planetary embryos has a constant value throughout the portion of the disk
under consideration. However, it is worth emphasizing that these conditions represent a
particular state of the disk. To consider that the solid mass in the inner region of the disk is 
equally distributed between embryos and planetesimals could be a reasonable assumption because in fact, this depends 
on the evolution of the system. However, considering that the population of planetesimals is homogeneously distributed throughout 
the inner region of the disk could not be. Planetesimal accretion rates are proportional 
to the surface density of solids and to the keplerian frequency. These two quantities increase closer to the star. 
Thus, it is expected that inner embryos grow quickly accreting all the available planetesimals. Due to the fact that 
embryos can only accrete planetesimals from their feeding zones, which are very narrow as we get closer to the central star, 
we expected a considerable number of small embryos, and practically no planetesimals in the inner region at the moment of the dissipation 
of the disk. 

Another consideration that could be not realistic is that all the embryos have the same mass in the inner 
region of the disk. Basically, the final mass of an embryo (during the gaseous phase) is a function of the distance to 
the star and of the time-scale dissipation of the disk. Brunini \& Benvenuto (2008) showed that this functional dependence is 
maximized near the snow line. Thus, it is expected that the most massive embryos of the distribution are located near the snow line.

In order to test the sensitivity of the results to the initial distributions of planetary embryos and planetesimals, we construct two
different sets of simulations from the same protoplanetary disk. The first one uses ad hoc initial conditions for
planetesimals and embryos like in \citet{RoncodeElia2014} while the second set assumes initial conditions derived from a
semi-analytical model, which analyzes in detail the evolution of embryos and planetesimals during the gaseous phase. 

Since our main goal is to analyze the formation process of terrestrial planets and to focus in those of the HZ, our region of study
extends between 0.5 au and 5 au. Our model assumes an {\it inner zone} in the disk between 0.5 au and the snow line, and an
{\it outer zone}, which extends between the snow line and 5 au.

In the following, we describe the generation of initial conditions for both sets of numerical simulations.

\begin{figure}[t]
 \centering
 \includegraphics[angle=270, width= 0.48\textwidth]{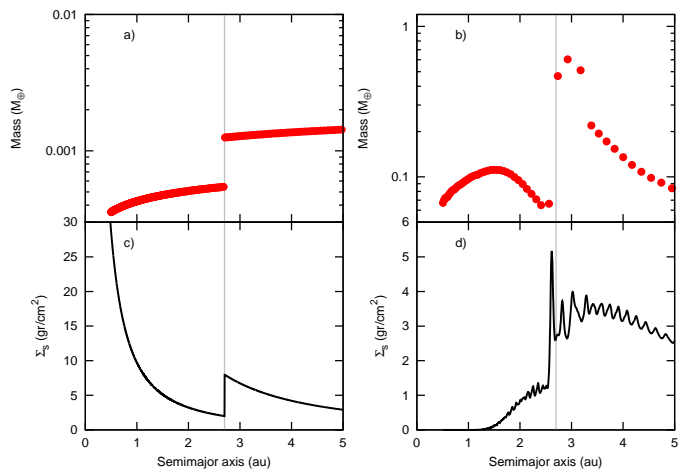}
 \caption{Initial and final distributions of embryos and planetesimals generated by the semi-analytical model. Figures a) and b) represent
   the initial and final mass distribution of embryos. At the beginning (a) there were 223 embryos between 0.5 au and 5 au, and after 3 Myr of 
evolution (b) there only remain 57 embryos. Figures c) and d) represent the initial and final density profile of planetesimals.
   Figures b) and d) then represent the final results obtained with the semi-analytical model after 3 Myr of evolution and the initial
   conditions for the N-body simulations. The vertical lines in grey represent the position of the snow line at 2.7 au. Color figure is
only available in the electronic version. }
 \label{fig:cond-ini}
\end{figure}

\subsection{First set: ad hoc initial conditions}

In this first set of simulations, we assumed the ad hoc initial conditions used in the work developed by \citet{RoncodeElia2014}.
Considering a protoplanetary disk of 0.03 $M_{\odot}$ and an exponent $\gamma$ = 1.5 associated with the surface density profile,
the total mass of solids in the region of study is 13.66 $M_{\oplus}$. In agreement with various planetary accretion studies such
as \citet{Kokubo1998}, this mass of solids is divided equally between the planetesimal and embryo components. 
On the one hand,
45 planetary embryos are distributed in the study region, 35 of which are in the inner zone with masses of 0.06 $M_{\oplus}$, and
10 in the outer zone with masses of 0.47 $M_{\oplus}$. On the other hand, 1000 planetesimals are distributed in the disk with
masses of 2.68 $\times$ 10$^{-3}$ $M_{\oplus}$ and 0.021 $M_{\oplus}$ in the inner and outer zone, respectively.

\subsection{Second set: initial conditions from a semi-analytical model}

In this second set of simulations, we generate initial conditions for embryos and planetesimals using a semi-analytical model based
on the works of \citet{BruniniBenvenuto2008} and \citet{Guilera2010} with the inclusion of some minor improvements. This model analyze
the evolution of a protoplanetary disk during the gaseous phase and thus, it is possible to obtain more realistic initial conditions to
be used in a N-body code. A detailed description of this model can be found in \citet{deElia2013} Sec.3.

In this semi-analytical model, the embryos are separated by 10 mutual Hill radii and the mass of
each embryo corresponds to the transition mass between runaway and oligarchic growth \citep{IdaMakino1993}. At the beginning there are
223 embryos in the study region, between 0.5 au and 5 au, which present a density of 3 gr cm$^{-3}$ and the planetesimals present a density of 1.5 gr cm$^{-3}$ and have radii of 10~km. The population of planetesimals is represented by the density profile of planetesimals.
In this model, the initial planetesimal size distribution evolves in time through planetesimal migration and
accretion by embryos. Embryos then grow by accretion of planetesimals and also by collisions between them.
We do not include type I migration in our model and the main reason for neglecting
this effect is that many quantitative aspects of the type I migration are still uncertain as we have already mentioned in Sec. 1. Besides, although planetesimals are affected by the gas drag, as they are so big (10 km), this migration is not important to produce significant changes in the final amounts of water on embryos and in the distribution of water along the planetesimal population. We consider that when the distance between
two embryos becomes smaller than 3.5 mutual Hill radii, they merge into one object. We assume perfect inelastic collisions.
For mergers between embryos, we neglected the presence of embryo gaseous envelopes because the masses of the embryos
are small enough not to produce major differences.

The final configuration obtained after the gas is completely dissipated in 3~Myr is also in agreement with that used \emph{ad hoc} 
in several works as mentioned in Sec. 3.1: we found the same proportion for both populations, half the mass in embryos and half 
the mass in planetesimals.
Figure \ref{fig:cond-ini} shows the results of our semi-analytical model for
the distribution of embryos and planetesimals for a disk of $0.03M_\odot$ at the end of the gas phase.
In particular, the top left panel represent the mass distribution of planetary embryos as a function of the distance from the central
star used to initiate the evolution of the gas phase with the semi-analytical model, while the left bottom panel shows the surface density
profile of planetesimals used to do the same. The top right panel represents the final distribution of embryos obtained with the
semi-analytical model after 3~Myr of evolution until the gas is already gone and the bottom right panel represents the final surface
density profile of planetesimals. These (right panels) final distributions of embryos and planetesimals represent the initial conditions
to be used in our N-body simulations. In this case, 57 planetary embryos are distributed between 0.5 au and 5 au and 1000
planetesimals with individual masses of $7.24 \times 10^{-3} M_{\oplus}$ are generated using the distribution shown in Fig. 1d).

\begin{figure}[ht]
 \centering
 \includegraphics[angle=0, width= 0.4\textwidth]{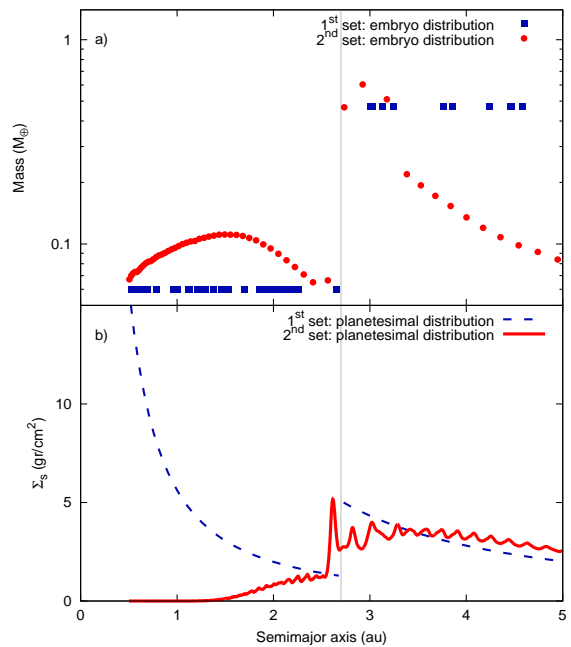}
\caption{a) Distributions of embryos used to start the N-body simulations. The squares represent the distribution of embryos obtained
with the first set of simulations and the circles represent the final results obtained with the semi-analytical model, which represent
the second set of simulations. b) Surface density profiles used to distribute 1000 planetesimals to start the N-body simulations. The
dashed line represents the surface density for the first set and the solid line represents the surface density for the second set
obtained with the semi-analytical model. Color figure is only available in the electronic version.}
 \label{fig:Distribuciones-CI1-CI3}
\end{figure}

\section{N-body simulations: characteristics}

Our N-body simulations begin when the gas of the disk has already dissipated.
The numerical code used in our N-body simulations is the MERCURY code developed by \citet{Chambers1999}. We particularly
adopted the hybrid integrator, which uses a second-order mixed variable simplectic algorithm to treat the interaction between
objects with separations greater than 3 Hill radii, and a Burlisch-Stoer method for resolving closer encounters.

In both sets of simulations, all collisions were treated as inelastic mergers
that conserve mass and water content. Given that the N-body simulations are very costly numerically, we decided to reduce the CPU
time considering only gravitational interactions between embryos and planetesimals \citep{Raymond2006}. Thus, planetesimals 
do not gravitationally interact and neither collide with each other. These N-body high-resolution simulations allowed us to describe in 
detail the dynamical processes involved during the formation and post evolution stages.

Since terrestrial planets in our Solar System might have formed in 100~Myr - 200~Myr \citep{Touboul2007, Dauphas2011},
we integrated each simulation for 200~Myr. To compute the inner orbit with enough precision
we used a 6 day time step. Moreover, the Sun's radius was artificially increased to 0.1 au in order to avoid numerical
error for small-perihelion orbits.

The orbital parameters, such as initial eccentricities and inclinations, were taken randomly considering values lower than
$0.02$ and $0.5^{\circ}$, respectively, both for embryos and planetesimals. In the same way, we adopted random values for the argument
of pericenter $\omega$, longitude of ascending node $\Omega$ and the mean anomaly $M$ between $0^\circ$ and $360^\circ$.

Due to the stochastic nature of the accretion process, we performed three simulations S1, S2 and S3 
with different random number seeds for every set of
initial conditions. It is worth mentioning that each simulation conserved energy to at least one part in $10^3$.

\section{Results}
  
\subsection{General comparative analysis}

The initial distributions associated to planetary embryos and planetesimals for both sets 
of simulations used as initial conditions for the N-body simulations are shown in Figure \ref{fig:Distribuciones-CI1-CI3}.
The blue distributions correspond to the simulations performed by \citet{RoncodeElia2014} and the red ones correspond to the new set of
simulations performed from a semi-analytical model. Panel a) presents the initial distributions of embryos corresponding to the end of
the gas phase and panel b) presents the surface density profiles used to distribute 1000 planetesimals in the study region, also when
the gas is already dissipated. 

As we can see, the embryos inside the snow line of the second set of simulations present masses that range from 1 to 2 times the 
masses of the embryos inside the snow line of the first set. Something similar happens with the embryos beyond the snow line. In fact, 
the masses of the outermost embryos of the first set of simulations are at most 5 times greater 
than those of the second set. Another difference evidenced in panel b) is that the surface density of planetesimals of both
sets are quite different in the region inside the snow line since the surface density profile of planetesimals for the second set of
simulations is almost zero between 0.5~au and 2.7~au. This is because the planetesimals in this region were accreted by the embryos 
during the gaseous phase. Could these differences lead to significant changes in the final results of our simulations or not? 
A discussion on this issue will be addressed in Section 6.

\begin{figure*}[ht]
 \centering
 \includegraphics[angle=270, width= 0.99\textwidth]{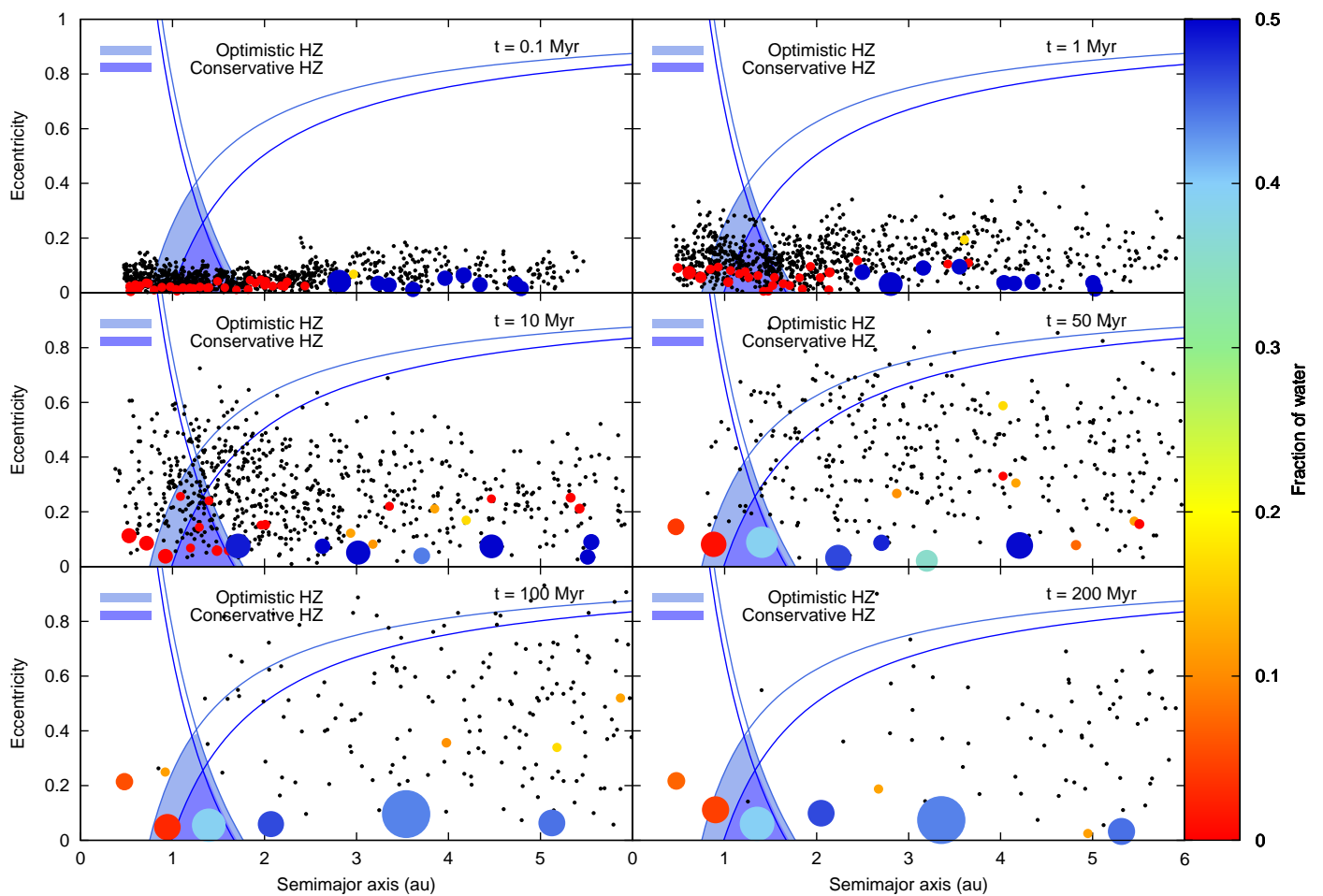}
 \caption{Evolution in time of the S2 simulation of the first set. The blue and light-blue shaded
areas represent the conservative and the optimistic HZ respectively and the blue and light-blue curves represent curves of constant
perihelion and aphelion, both for the conservative and the optimistic HZ. Planetary embryos are plotted as colored circles and
planetesimals are plotted with black dots. The color scale represents the fraction of water of the embryos respect to their total
masses. In this case there are two planets within the optimistic HZ with masses of $1.19M_\oplus$ and $1.65M_\oplus$, respectively.
They present $4.51\%$ and $39.48\%$ of water content by mass, which represents $192$ and $2326$ Earth's oceans, respectively.
Color figure is only available in the electronic version.}
 \label{fig:g15-S2-adhoc}
\end{figure*}

\begin{figure*}[ht]
 \centering
 \includegraphics[angle=270, width= 0.99\textwidth]{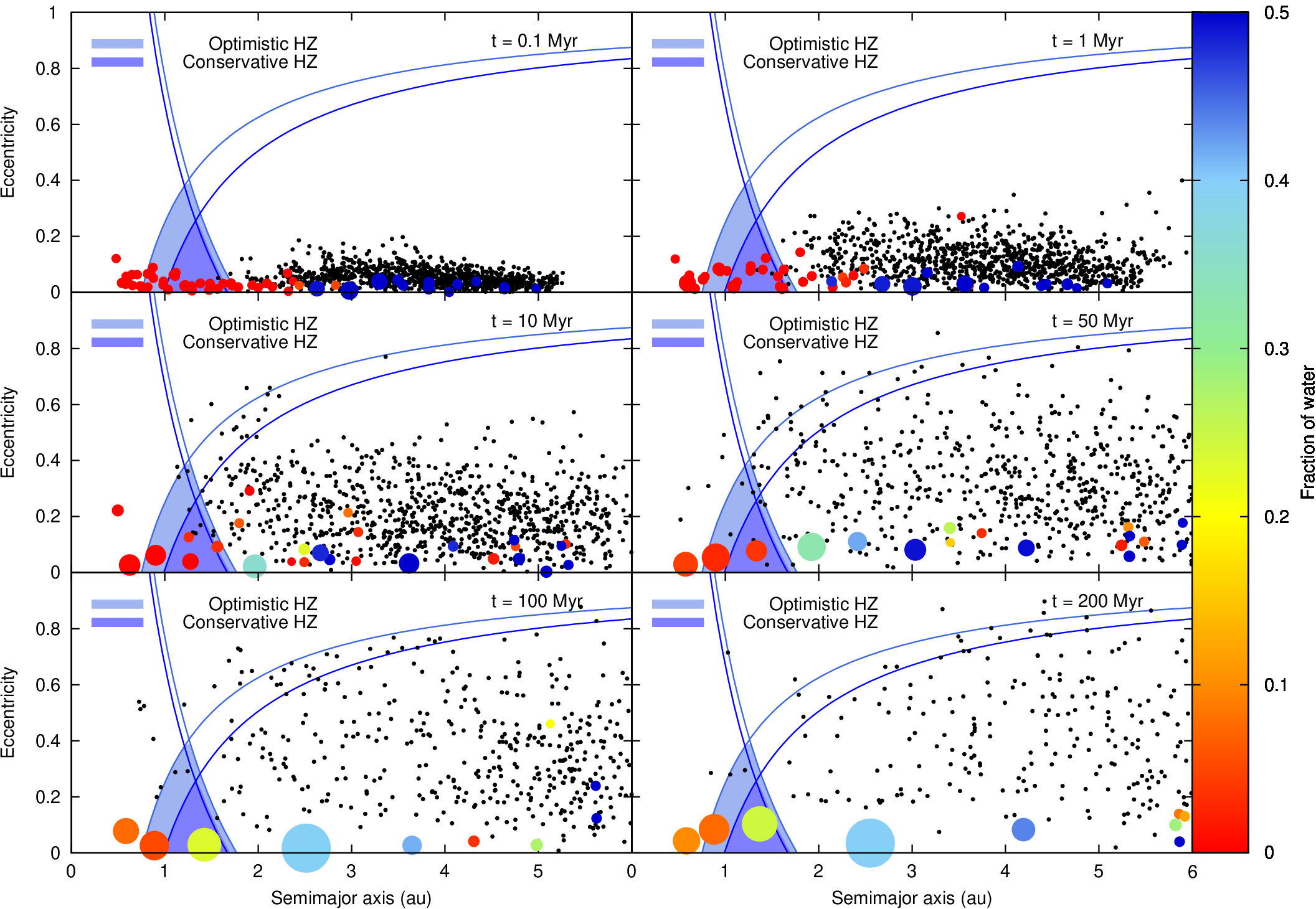}
 \caption{Evolution in time of the S3 simulation of the second set. The blue and light-blue shaded
areas represent the conservative and the optimistic HZ respectively and the blue and light-blue curves represent curves of
constant perihelion and aphelion, both for the conservative and the optimistic HZ. Planetary embryos are plotted as colored
circles and planetesimals are plotted with black dots. The color scale represents the fraction of water of the embryos respect
to their total masses. Two planets remain within the optimistic HZ with masses of $1.37M_\oplus$ and $1.65M_\oplus$,
respectively. They present $7.49\%$ and $24.38\%$ of water content by mass, which represents $366$ and $1437$ Earth's oceans,
respectively. Color figure is only available in the electronic version.}
 \label{fig:g15-S3-semi}
\end{figure*}

Figures \ref{fig:g15-S2-adhoc} and \ref{fig:g15-S3-semi} show six snapshots in time on the semimajor axis eccentricity plane of
the evolution of the S2 and S3 simulations of the first and the second set, respectively.
In general terms, the overall progression of both sets of simulations can be described as follows. From the beginning, the planetary
embryos are quickly excited by their own mutual gravitational perturbations. At the same time, the planetesimals, which are not
self-interacting bodies, significantly increase their eccentricities due to perturbations from embryos. In time, the eccentricities
of embryos and planetesimals increase until their orbits cross and accretion collisions occur. Then, planetary embryos grow by accretion
of other embryos and planetesimals and the total number of bodies decreases. Finally, at the end of the simulations, the entire study
region from 0.5~au and 5~au contains between four and seven planets with separations ranging from 18 to 53 mutual hill radii and
a total mass that represents between $55\%$ and $56\%$ of the initial mass of solids in the disk which is $13.66M_\oplus$.

\begin{figure}[t]
 \centering
 \includegraphics[angle=0, width= 0.5\textwidth]{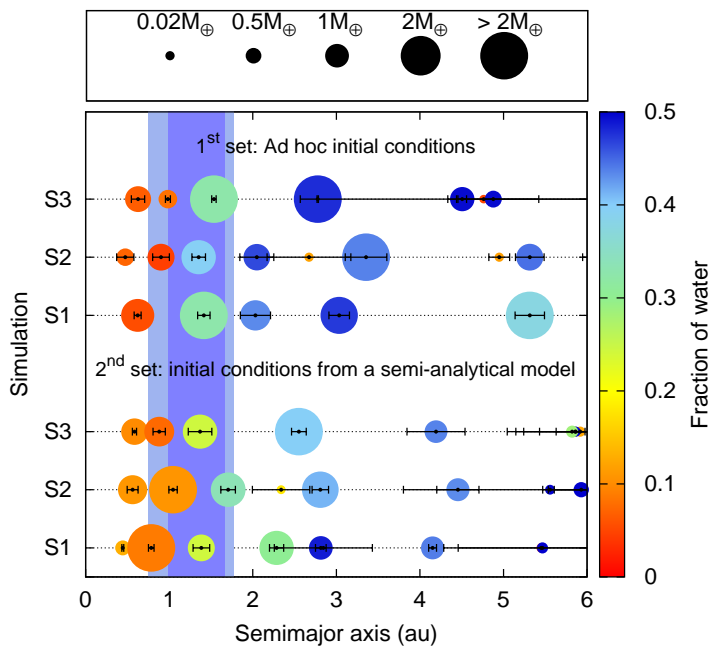}
\caption{Final configuration of the first and second set of  simulations. The
color scale represent the water content and the shaded regions, the optimistic and the conservative HZ. The eccentricity of each
planet is shown over it, by its radial movement over an orbit. Color figure only available in the
electronic version.} 
 \label{fig:g15-comparacion}
\end{figure}

Between the resulting planets of each simulation, it is possible to distinguish three classes of special interest: 1) the innermost
planet of the system; 2) the planet (or planets) in the HZ; and 3) the most massive planet of the system. Tables \ref{tab:1} and
\ref{tab:2} show the general characteristics of these three distinctive planets resulting from S1, S2, and S3 simulations for the
first and the second set of initial conditions, respectively, and Fig. \ref{fig:g15-comparacion} shows the final configuration 
of all the simulations performed. The main similarities and differences obtained in these simulations
for the resulting planets as well as the remaining planetesimal population are discussed throughout the following subsections.

\begin{table*}[t!]
\caption{General characteristics of the most distinctive planets formed in the S1, S2, and S3 simulations of the first set,
using ad hoc initial conditions. $a_{\text{i}}$ is the initial and $a_{\text{f}}$ is the final semimajor axis. The initial semimajor axis of a planet represents the initial location of the most massive embryo from which the planet formed.
$<e>_{100\text{Myr}}$ and $<i>_{100\text{Myr}}$ represent the mean eccentricity and inclination, respectively, in the last
100~Myr of evolution. $M$($M_\oplus$) represents the final mass, $W$($\%$) the final percentage of water by mass and
$T_{\text{LGI}}$ the timescale in Myr associated with the last giant impact produced by an embryo.}
\begin{center}
\begin{tabular}{|c|c|c|c|c|c|c|c|c|}
\hline 
\hline
Simulation  & Planet          & $a_{\text{i}}$(au) & $a_{\text{f}}$(au) & $<e>_{100\text{Myr}}$ & $<i>_{100\text{Myr}}$ & $M$($M_\oplus$)  & $W$($\%$) & $T_{\text{LGI}}$ (Myr) \\
\hline
\hline
S1          &  Innermost                & 0.68             & 0.62           & 0.07 & $4^\circ.25$ & 1.57 & 5.47   & 56.1  \\
            &  Planet \emph{a} in HZ    & 4.52             & 1.41           & 0.06 & $2^\circ.39$ & 2.21 & 32.55  & 35    \\
            &  Most massive             & 4.52             & 1.41           & 0.06 & $2^\circ.39$ & 2.21 & 32.55  & 35    \\
\hline
S2          &  Innermost                & 0.78             & 0.47           & 0.14 & $5^\circ.86$ & 0.6  & 7.12   & 5.55  \\     
            &  Planet \emph{b} in HZ    & 1.30             & 0.90           & 0.08 & $3^\circ.56$ & 1.19 & 4.51   & 35    \\
            &  Planet \emph{c} in HZ    & 3.13             & 1.35           & 0.06 & $2^\circ.86$ & 1.65 & 39.48  & 22    \\
            &  Most massive             & 3.25             & 3.35           & 0.05 & $2^\circ.20$ & 2.15 & 43.66  & 96.8  \\
\hline
S3          &  Innermost                & 0.50             & 0.63           & 0.08 & $4^\circ.04$ & 1.13 & 6.66   & 11.45 \\      
            &  Planet \emph{d} in HZ    & 0.64             & 0.98           & 0.07 & $3^\circ.64$ & 0.66 & 8.10   & 6.00  \\
            &  Planet \emph{e} in HZ    & 3.43             & 1.53           & 0.05 & $2^\circ.23$ & 2.27 & 32.46  & 71.9  \\
            &  Most massive             & 4.69             & 2.77           & 0.04 & $1^\circ.47$ & 3.08 & 47.86  & 44.3  \\
\hline
\hline
\end{tabular}
\end{center}
\label{tab:1}
\end{table*}

\begin{table*}[t!]
\caption{General characteristics of the most distinctive planets formed in the S1, S2, and S3 simulations of the second set, using initial
conditions from a semi-analytical model. $a_{\text{i}}$ is the initial and $a_{\text{f}}$ is the final semimajor axis. The initial semimajor axis of a planet represents the initial location of the most massive embryo from which the planet formed. $<e>_{100\text{Myr}}$ and $<i>_{100\text{Myr}}$ represent the mean eccentricity and inclination, respectively, in the last 100~Myr of
evolution. $M$($M_\oplus$) represents the final mass, $W$($\%$) the final percentage of water by mass and $T_{\text{LGI}}$ the
timescale in Myr associated with the last giant impact produced by an embryo.}
\begin{center}
\begin{tabular}{|c|c|c|c|c|c|c|c|c|}
\hline 
\hline
Simulation  & Planet          & $a_{\text{i}}$(au) & $a_{\text{f}}$(au) & $<e>_{100\text{Myr}}$ & $<i>_{100\text{Myr}}$ & $M$($M_\oplus$)  & $W$($\%$) & $T_{\text{LGI}}$ (Myr) \\
\hline
\hline
S1          &  Innermost                & 0.54           & 0.44           & 0.12 & $8^\circ.76$ & 0.45 & 13.03  & 74.24  \\
            &  Planet \emph{a} in HZ    & 1.05           & 0.78           & 0.05 & $2^\circ.36$ & 2.21 & 8.75   & 51.6  \\
            &  Planet \emph{b} in HZ    & 1.49           & 1.38           & 0.06 & $3^\circ.02$ & 1.18 & 24.30  & 28.2  \\ 
            &  Most massive             & 1.05           & 0.78           & 0.05 & $2^\circ.36$ & 2.21 & 8.75   & 51.6  \\
\hline
S2          &  Innermost                & 0.87           & 0.56           & 0.07 & $3^\circ.93$ & 1.33 & 10.93  & 198.8 \\
            &  Planet \emph{c} in HZ    & 0.73           & 1.04           & 0.04 & $2^\circ.01$ & 2.00 & 10.76  & 20.5  \\ 
            &  Planet \emph{d} in HZ    & 2.73           & 1.70           & 0.05 & $2^\circ.09$ & 1.67 & 33.63  & 90.5  \\ 
            &  Most massive             & 0.73           & 1.04           & 0.04 & $2^\circ.01$ & 2.00 & 10.76  & 20.5  \\
\hline
S3          &  Innermost               & 0.87           & 0.58           & 0.06 & $2^\circ.47$ & 1.18 & 9.90   & 34.8  \\
            &  Planet \emph{e} in HZ   & 1.33           & 0.88           & 0.06 & $2^\circ.09$ & 1.37 & 7.49   & 18.3  \\
            &  Planet \emph{f} in HZ   & 1.14           & 1.37           & 0.05 & $1^\circ.67$ & 1.65 & 24.38  & 59.7  \\
            &  Most massive            & 2.92           & 2.55           & 0.03 & $1^\circ.55$ & 2.36 & 39.74  & 62.7  \\
\hline
\hline
\end{tabular}
\end{center}
\label{tab:2}
\end{table*}

\subsection{Removal of embryos and planetesimals}

It is worth mentioning that while the total mass of the planetesimal population starting the N-body simulations is almost the
same for both sets of initial conditions ($\sim 6.9M_\oplus$ and $7.2M_\oplus$ for the first and the second set, respectively), 
the initial surface density profile of planetesimals is very different in both
cases (see Fig. \ref{fig:Distribuciones-CI1-CI3}b). In fact, for the first set of simulations, 31\% of the mass of the planetesimal 
population is located inside the snow line, while the remaining 69\% is distributed beyond 2.7 au. On the other hand, for the second 
set of initial conditions, only 10\% of the mass of the planetesimal population is distributed inside the snow line, while the 
remaining 90\% is located beyond 2.7 au. Thus, some differences should be observed in the evolution of the planetesimal population 
between both sets of initial conditions.

After 200 Myr of evolution, many embryos and planetesimals were removed from the disks of both sets of simulations.
In the S2 simulation of the first set of initial conditions, the mass of planetary embryos and planetesimals that still
remain in the study region, between 0.5 au and 5 au, is $\sim 7M_\oplus$ and 0.43$M_\oplus$, respectively. We consider that this 
remaining planetesimal mass, which represents $6.3\%$ of the initial planetesimal mass, is not significant
to produce important changes in the orbital and physical characteristics of the resulting planets. Similar results were
found in the other simulations of this set.

In the S3 simulation of the second set of initial conditions, the mass that still survive after 200 Myr is 7.49$M_\oplus$ in embryos 
and 1.08$M_\oplus$ in planetesimals. Thus, the remaining mass of planetary embryos in the study region is comparable in both sets of initial
conditions. However, the mass of remaining planetesimals associated to the second set of simulations, which represents $15\%$ of the 
initial planetesimal mass, is a factor of $\sim$ 2.4 greater than that obtained from the first set of initial conditions.
This situation is repeated in the rest of the simulations of this set.

From a quantitative point of view, the populations of embryos at the end of a simulation obtained
from both sets of initial conditions result to be comparable, while the remaining populations of planetesimals show the main discrepancies.
In order to understand such differences, it is very important to analyze the main mechanisms of removal of planetesimals in our
simulations. On the one hand, in the S2 simulation of the first set, $3.78M_\oplus$ in planetesimals, which represents $55\%$ 
of the initial mass in planetesimals, collide with the central star, remains outside the study region (beyond 5~au) or is ejected 
from the disk. On the other hand, in the S3 simulation of the second set $3.83M_\oplus$, which represents $53\%$ of the initial mass 
in planetesimals, collide with the central star, remains outside the study region or is ejected from the disk. Then, the removed mass 
in planetesimals is also comparable in both sets of 
simulations. However, the accreted mass in planetesimals (by embryos) represents $37.9\%$ of the initial mass in planetesimals for the S2 
simulation of the first set and represents $32.3\%$ of the initial mass in planetesimals for the S3 simulation of the second set. 
Similar results were found for the rest of the simulations. While the removed mass in planetesimals is comparable in both sets 
and range between $\sim 53\%$ and $55\%$ of the initial planetesimal mass for the second and the first set, the percentage of the 
accreted mass in planetesimals range between $37.9\%$ and $41\%$ for the first set and between $31.7\%$ and $32.3\%$ for the second set. This 
difference in the accreted mass in planetesimals between both sets of simulations reveals distinct accretion histories of the 
final planets. The accretion history of the planets is of great interest because the final characteristics and properties of the
planets depend on it. Thus, initial conditions obtained from more realistic models allow us to obtain more reliable physical 
properties of the resulting planets like, for example, a more reliable final water distribution.

\subsection{Dynamical friction}

\begin{figure}[ht]
 \centering
 \includegraphics[angle=270, width= 0.47\textwidth]{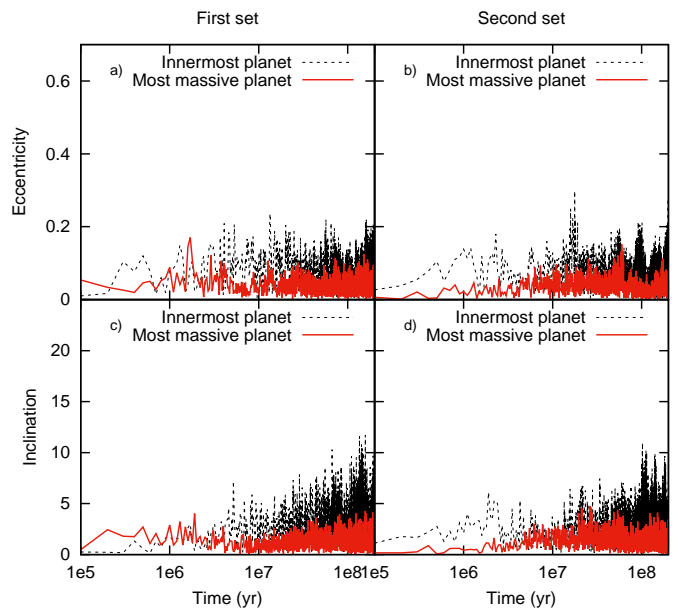}
\caption{Evolutions of eccentricities, a), b), and inclinations, c), d), as a function of time for the innermost planet 
(black curve) and the most massive planet (red curve) of the S3 simulations and S3 simulations in the first and in the second 
set, respectively. The dynamical friction effects prevail over the larger bodies which evolve, in both sets of simulations, 
in regions of the disk embebbed in a swarm of planetesimals. Color figure is only available in the electronic version.}
 \label{fig:Fd-comparacion}
\end{figure}

From the beginning, the dynamical friction seems to be a significant phenomenon in both sets of simulations. 
This dissipative force dampens the eccentricities and inclinations of the large bodies embedded in a swarm of smaller bodies
represented by planetesimals. Figure \ref{fig:Fd-comparacion} shows the evolution of the eccentricities and inclinations of the 
innermost planet and the most massive planet in S3 of the first set and in S3 of the second set. The masses of the innermost planets 
are comparable and the same happens with the most massive planets (see Tables \ref{tab:1} and \ref{tab:2}).
Thus, we find similar results for both of them. 
 
On the one hand, the innermost planets of $1.13M_{\oplus}$ and $1.18M_{\oplus}$ show mean values of the eccentricity of
0.08 and 0.06 for the first and the second set, respectively, while the mean values for the inclinations are $4.04^{\circ}$ and 
$2.47^{\circ}$. On the other hand, the most massive planets of $3.08M_{\oplus}$ and $2.36M_{\oplus}$ present mean values of the
eccentricity of 0.04 and 0.03 for the first and the second set, respectively, while the mean values for the inclinations are 
$1.47^{\circ}$ and $1.55^{\circ}$. This means that, for both kinds of simulations, the effects of dynamical friction still prevail over
larger bodies. It is worth noting that both planets are initially located in regions embedded in planetesimals, therefore, it is 
expected to see the same effects in both of them.

Similar results are obtained with the comparison between S2 of the first set and S1 of the second set. In this particular scenario, 
it is worth noting that the most massive planet in S1 of the second set is not located in a region embedded in planetesimals at the 
beginning of the integration. However, this planet has of the order of $5 \times 10^{5}$ close encounters with planetesimals
during its evolution, which allow to keep the low values associated to its eccentricity and inclination. Therefore, 
the fact that the planet is located at the beginning in a region without planetesimals does not implies that the effect of dynamic friction is absent throughout its evolution.

An interesting scenario to analyze is represented for the S2 simulation of the second set. In this particular case, the innermost 
and the most massive planets started the simulation in a region of the disk which is not embedded in planetesimals at the beginning 
of the integration. It is important to note that the final masses of such planets have comparable values. However, the mean values 
of the eccentricity and inclination of the innermost planet are a factor of $\sim$ 2 greater than those associated to
the most massive planet. Such a difference is due to the fact that the most massive planet suffered one order of magnitude
more close encounters with planetesimals that the innermost planet throughout the evolution.
Thus, the results of our simulations suggest that the dynamical friction is an inefficient mechanism for the innermost planets.

\subsection{Planets in the habitable zone}

The formation of terrestrial planets in the HZ of the system and the analysis of their water contents are topics of special
interest in our work. The HZ is defined as the range of heliocentric distances at which a planet can retain liquid water on
its surface \citep{Kasting1993}. For the case of the Solar System, \citet{Kopparapuerrata2013, Kopparapu2013} proposed that an 
\emph{optimistic} HZ is between 0.75~au and 1.77~au and a \emph{conservative} HZ is between 0.99~au and 1.67~au. However, it is 
worth noting that the fact that a planet is in the HZ does not guarantee that there may be developed life. Planets with very 
eccentric orbits may pass most of their periods outside the HZ, not allowing long times of permanent liquid water on their surfaces.
To avoid this problem, we considered that a planet is in the optimistic HZ if it has a perihelion $q \ge 0.75$~au and a
aphelion $Q \le 1.77$~au and is in the conservative HZ if it has a perihelion $q \ge 0.99$~au and a aphelion $Q \le 1.67$~au.
In this work, we considered it is sufficient that a planet is in the optimistic HZ to be potentially habitable. On the other
hand, we considered the water contents to be significant when they are similar to that of the Earth.
The mass of water on the Earth surface is 2.8 $\times$ 10$^{-4}$ $M_{\oplus}$, which is defined as 1 Earth ocean. Taking into
account the water content in the Earth mantle, several studies \citep{Lecuyer1998, Marty2012} suggested that the
present-day water content on Earth should be $\sim 0.1\%$ to $0.2\%$ by mass, which represents between $\sim 3.6$ to $7.1$ Earth oceans.
Moreover, some works suggest that the primitive mantle of the Earth could have contained between 10 and 50 Earth oceans 
\citep{Abe2000,RighterDrake1999}, although this is still under debate. 

We developed three different N-body simulations for every of the two sets of initial conditions. It is worth noting that all of them 
form planets within the optimistic HZ. For the first set of initial conditions, our simulations produce a total
of five planets in the HZ with masses from 0.66 $M_\oplus$ to 2.27 $M_\oplus$ and final water contents ranging from 4.5\% to
39.48\% of the total mass, which represents from 192 to 2326 Earth oceans. For the second set, our simulations form a total number
of six planets in the HZ with masses ranging from 1.18 $M_\oplus$ to 2.21 $M_\oplus$ and final water contents from 7.5\% to 33.63\%
of the total mass, which represents from 427 to 2006 Earth oceans.

\begin{figure*}[t!]
 \centering
 \includegraphics[angle=270, width= 0.8\textwidth]{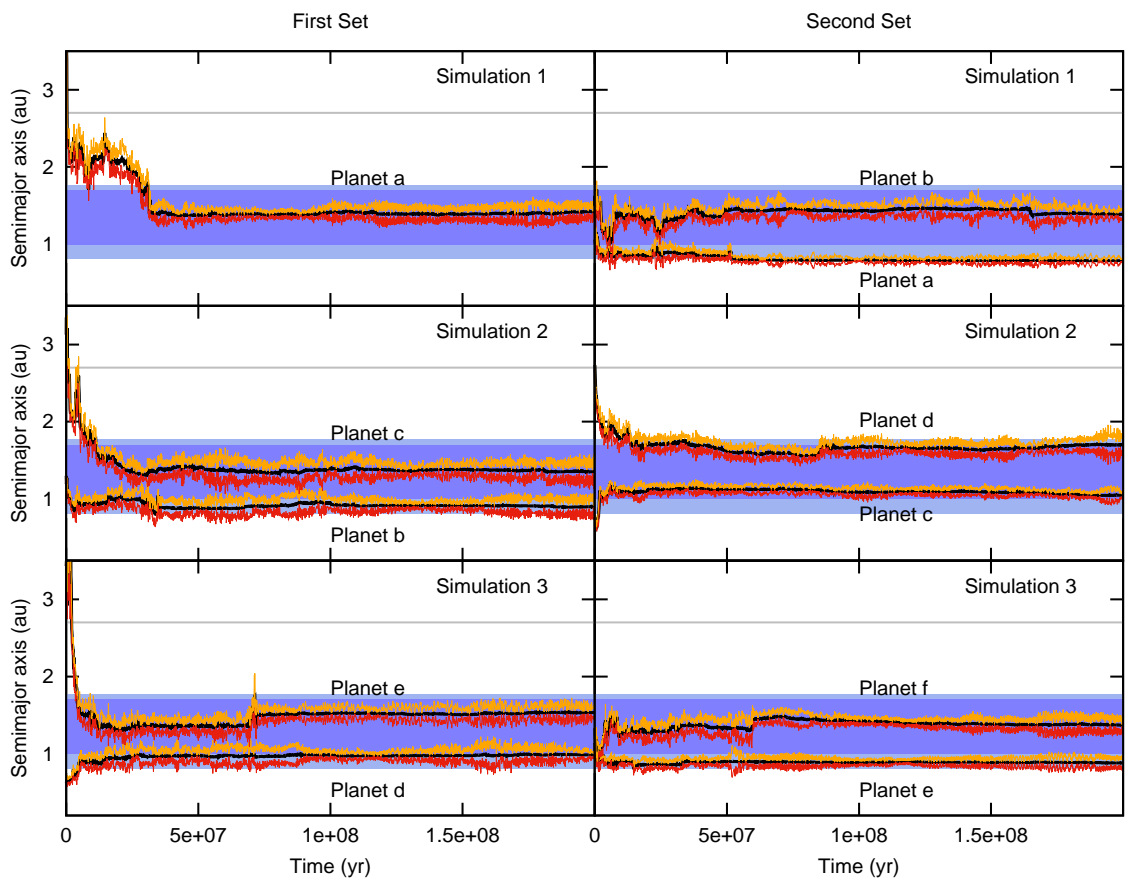}
 \caption{Evolution in time of the semimajor axis (black), the perihelion (red) and the aphelion (orange) of the planets that remain
 within the HZ of each simulation in both, the first (left panel) and the second set (right panel) of simulations. The blue and light-blue 
shaded areas represent the conservative and the optimistic HZ respectively and the dashed grey line represents the position of the snow 
line. Color figure is only available in the electronic version.}
 \label{fig:3}
\end{figure*}

Figure \ref{fig:3} shows the evolution of the semimajor axis, the perihelion and the aphelion of the planets that remain in
the HZ at the end of the simulation for both sets of initial conditions. As we can see, most of the planets are kept within the
conservative HZ almost throughout the whole integration, and their semimajor axes do not change significantly in the last 100 Myr
of evolution. Planet \emph{b} (S2 simulation) in the first set of initial conditions as well as planets \emph{a} (S1 simulation)
and \emph{e} (S3 simulation) in the second set remain in the limits of the optimistic HZ, with perihelions entering and leaving it
due to oscillations in the eccentricity of their orbits. Since the definition of HZ is not accurate, we will consider that these
planets are targets of potential interest. Planet \emph{d} (S2 simulation) in the second set of initial conditions
reaches maximum values of the eccentricity of $\sim$ 0.15. This planet shows changes in its aphelion, which carry it in and out of the
HZ at the end of the simulation. The incident stellar flux on such a planet may have considerable variations between perihelion and 
aphelion. The study developed by \citet{WilliamsPollard2002} shows that, provided that an ocean is present 
to act as a heat capacitor, it is primarily the time-averaged flux that affects the habitability over an eccentric orbit. Given that 
planets on eccentric orbits have higher average orbital flux, planet \emph{d} (S2 simulation) in the second set may maintain habitable 
conditions.

\subsubsection{Highly water-rich planets}

As the reader can see in Tables 1 and 2, simulations developed from both sets of initial conditions form planets in the HZ with a
wide range of final water contents. In particular, some of them show very high water contents. In fact, planets
\emph{a}, \emph{c} and \emph{e} in the first set of simulations show water contents of 32.6\%, 39.5\% and, 32.5\% by mass,
respectively, while planets \emph{b}, \emph{d} and \emph{f} in the second set present 24.3\%, 33.6\%, and 24.4\% water by
mass, respectively. It is worth noting that the evolution of such water-rich planets has different characteristics for both sets of
simulations. Tables 1 and 2, and Figure \ref{fig:3} give us an interesting vision concerning this point. In fact, Table 1 and
Figure \ref{fig:3} (left panel) show us that potentially habitable planets \emph{a}, \emph{c} and \emph{e} of the first set of
simulations come from beyond the snow line. The migration suffered by these planets is due to the strong gravitational interaction
between planetary embryos and planetesimals in the outer disk. Given their initial locations, it is worth noting that an important
percentage (31\%-36\%) of the very high final water contents of such planets results to be primordial. Moreover, our simulations
indicate that the remaining water content of those planets is acquired by impacts of planetesimals and planetary embryos in equal
parts. Thus, taking into account the primordial and acquired water contents, we infer that the embryos play the most important role to
provide the water to the potentially habitable planets \emph{a}, \emph{c} and \emph{e} of the first set of simulations.

Table 2 and Figure \ref{fig:3} (right panel) show us that potentially habitable planets \emph{b}, \emph{d} and \emph{f} of the second set
of simulations present different characteristics. On the one hand, planet \emph{d} starts the simulation beyond the snow line, like
the potentially habitable planets \emph{a}, \emph{c} and \emph{e} of the first set. In the same way, 41.4\% of the high final water
content of this planet is primordial, being a consequence of its initial location. Moreover, the remaining water content is provided
equally by impacts of embryos and planetesimals associated to the outer disk. Thus, taking into account the primordial water content,
we find that the embryos are the main responsible to provide the water to planet \emph{d} of the second set of simulations.
On the other hand, planets \emph{b} and \emph{f}
present a semimajor axis which do not change significantly from the beginning of the integration. This means that such planets grow
near their initial positions and do not suffer significant migration. Given their initial locations, the primordial water content
of the planets \emph{b} and \emph{f} is negligible. In fact, for both planets, the final water content is fully acquired throughout
evolution by impacts of planetesimals and planetary embryos associated to the outer disk. This can be seen in the bottom panel of 
Fig. \ref{fig:4} where the feeding zones of the planets are represented. In particular, planets \emph{b} and \emph{f} accreted two and one 
embryos from beyond the snow line respectively, which significantly increased their final water contents. However, planetesimals provide
53\% and 44\% of the final water contents of planet \emph{b} and \emph{f}. Thus, for these cases, both embryos and 
planetesimals are equally responsible to provide the final water contents of such planets.   

Although these highly water-rich planets remain within the optimistic habitable zone, their 
potential habitability is still under debate \citep{Abbot2012,Alibert2014,Kitzmann2015} 
and that analysis is beyond the scope of this work. However we consider water worlds as a particular and very interesting 
kind of exoplanets.

\subsubsection{Earth-like planets}

\begin{figure}[ht]
 \centering
 \includegraphics[angle=0, width= 0.45\textwidth]{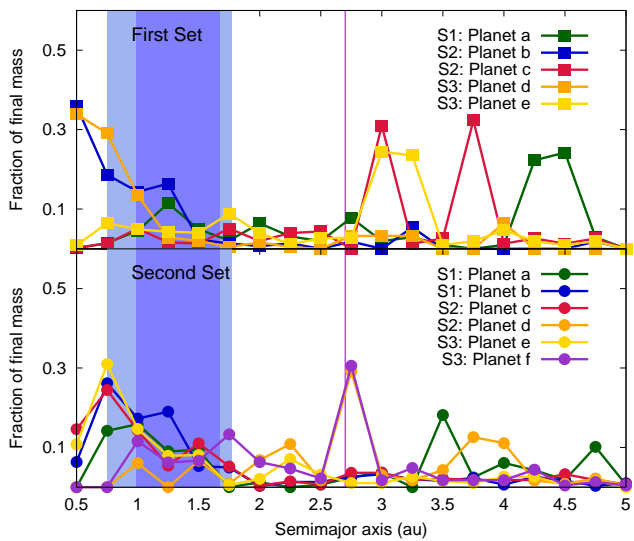}
 \caption{Feeding zones of the planets that remain in the HZ of S1, S2 and S3 simulations of the first (top panel) and the second 
 set of simulations (bottom panel). The y axis represents the fraction of each planet's final mass after 200 Myr of evolution. Color figure
is only available in the electronic version.}
 \label{fig:4}
\end{figure}

Our simulations also form \emph{Earth-like} or \emph{terrestrial-type} planets within the optimistic HZ. 
We consider the Earth is comparable to these planets not only because they present masses similar to that of the
Earth, but rather because of the features that have to do with the dynamics of their formation. These Earth-like planets 
in our simulations form \emph{in situ}. There is no significant migration of the accretion seeds that end up forming
such planets. Moreover, their feeding zones, unlike what happens with highly water-rich planets, are restricted 
to the inner region of the disk (see Fig. \ref{fig:4}). This kind of exoplanets results to be the most interesting 
one from an astrobiological point of view since, as they present similarities in terms of the dynamics of the 
formation of the Earth, they are more likely to develop life.
It is worth mentioning that these Earth-like planets formed in our simulations present several times the Earth's
water content. This is directly related to two important facts: on the one hand, as we mentioned before, our model does not 
consider water loss during impacts, thus, the final water contents represent upper limits. On the other
hand, our formation scenario without gas giants is particularly conducive to a more efficient water delivery from external planetesimals. 
Giant planets could act as a barrier preventing that many of the external planetesimals, which are now accreted by the planets in the HZ, 
reach the inner zone of the disk \citep{Raymond2009}.

Two of the five planets formed within the optimistic HZ in the first set and 
three of the six planets that remain in the HZ of the second set 
are Earth-like planets. The Earth-like planets in the first set (see Table 1: planet b in S2 and
planet d in S3) present masses of $1.19M_\oplus$ and $0.66M_\oplus$ and water contents of $4.51\%$ and
$8.10\%$ by mass, respectively, which represent 192 and 190 Earth oceans. The Earth-like planets
in the second set (see Table 2: planet a in S1, planet c in S2 and planet e in S3) present masses 
of $2.21M_\oplus$, $2M_\oplus$ and $1.37M_\oplus$, and water contents of $8.75\%$, $10.76\%$ and
$7.49\%$ by mass, respectively, which represent 692, 768 and 366 Earth oceans. 

The evolution of these Earth-like planets presents similar characteristics in both sets of simulations
unlike what happens with the water worlds already described. Figure \ref{fig:3} shows that planets b and d
in the first set, and planets a, c and e in the second set, start their formation in the 
inner zone of the disk, particularly inside 1.5 au, and they evolve without exceeding this limit. 
This means that the percentage of water that they have achieved during their evolution is not primordial.
In fact, a detailed analysis of their collisional histories reveals that planetesimals are the only 
population responsible for their final amounts of water. Moreover, none of them accreted embryos
from beyond the snowline. However there is a difference regarding what class of planetesimals these
planets have accreted. From now on we will call planetesimals \emph{class-D} to those planetesimals
which were originally located inside the snow line and are dry, and we will call planetesimals
\emph{class-W} to those planetesimals which were originally located beyond the snow line
and present half of their mass in water. Planets b and d of the first set accreted 185 and 100 
planetesimals of the total of planetesimals, of which only 5 were class-W. This 
represents only $2.7\%$ and $5\%$ of the total amount of accreted planetesimals. In contrast, 
planets a, c and e of the second set accreted 65, 69 and 41 planetesimals of the total accreted planetesimals, 
of which 53, 59 and 28 were class-W planetesimals. This represents
$81.5\%$, $85.5\%$ and $68\%$ of the total amount of accreted planetesimals for planets a, c and e, respectively. 
Thus, the class-W planetesimals accreted for both 
sets of simulations is quite difference. Nevertheless it is necessary to remember that the masses of
the class-W planetesimals are different in both sets of simulations. We can note that 
the mass of water accreted by the Earth-like planets considering only class-W planetesimals in the 
second set is between 1.9 and 4 times the mass of water accreted by the planets of the first set.

An important topic to consider is the comparison between the timescale for an Earth-like planet to accrete 
half of the class-W population of planetesimals ($t_{\text{50.class-W}}$), the timescale to accrete half of the 
class-D planetesimals ($t_{\text{50.class-D}}$) and the timescale for the same planet to achieve at least $50\%$ 
of its final mass ($T_{\text{50\%}}$). Table \ref{tab:3} shows the main characteristics of the Earth-like planets 
formed within the HZ in both sets of simulations and shows that, for all of them, $t_{\text{50.class-W}}$ is 
greater than $t_{\text{50.class-D}}$, greater than $T_{\text{50\%}}$ and even more, is greater than $T_{\text{75\%}}$. 
This means that class-W population of planetesimals was accreted significantly later than class-D planetesimals 
during the formation process and after the planet achieved half of its final mass. Moreover, class-W 
planetesimals were accreted after the planet achieved $75\%$ of its final mass. This late arrival of the 
class-W planetesimals during the formation of terrestrial planets may favor water retention on the 
surface of the planet brought by these planetesimals. \citet{OBrien2014} studied the water delivery and giant 
impacts on terrestrial planets in the ``Grand Tack'' scenario. Particularly, \citet{OBrien2014} analyzed the water 
delivery of primitive planetesimals (``C-Type'' planetesimals in \citealp{Walsh2011}) which were initially located 
between the orbits of Jupiter and Saturn, or beyond. Although the Grand Tack scenario is quite different from 
the one we proposed since our planetary systems do not include giant planets, we find similar results 
concerning timescales of primitive planetesimals accretion.

Moreover, we find that the Earth-like planets of the first and the second set are still 
accreting water due to class-W planetesimals after $\sim$ 90 Myr of evolution. 
Although planets b and d of the first set
only accreted 5 class-W planetesimals, 3 of them were accreted after 90 Myr, particularly between 110 and 170 Myr. 
These planetesimals provided $58.7\%$ and $39.3\%$ of the total accreted water by planets b and d, 
respectively, which represent $2.64\%$ and $3.2\%$ of their final masses. For the second set, planet a
accreted 10 class-W planetesimals between 93 and 191 Myr which provided $18.7\%$ of the total 
accreted water, and represent $1.64\%$ of its final mass. Planet c accreted 20 class-W planetesimals 
between 91 and 199 Myr which provided $33.6\%$ of the total accreted water and represent $3.62\%$ of its final mass
and finally planet e accreted 11 class-W planetesimals between 92 and 192 Myr which provided $38.8\%$ 
of the total accreted water and represent $2.91\%$ of its final mass. This analysis may be of particular
interest for the habitable zone planets formed around M dwarfs regarding the work of \citet{TianIda2015}. 
Although the Earth-mass planets found by \citet{TianIda2015} in M dwarf habitable zones may have 
lost their water content through dissociation and hydrodynamic scape during the first $\sim$ 90 Myr of evolution, 
they could still accrete ice planetesimals and retain significant amounts of water on their surfaces.

\begin{table*}[t!]
\caption{General characteristics of the Earth-like planets formed within the HZ in both sets of simulations. $M$($M_\oplus$) 
represents the final mass, $t_{\text{50.class-D}}$ represents the timescale in Myr for a planet to accrete half of the class-D 
planetesimals that will accrete, $t_{\text{50.class-W}}$ represents the timescale in Myr for a planet to accrete half of the class-W 
planetesimals that will accrete, $T_{\text{50\%}}$ represents the timescale in Myr for a planet to achieve $50\%$ of its final mass and 
$T_{\text{75\%}}$ represents the timescale in Myr for a planet to achieve $75\%$ of its final. Planet b of the first set shows the same 
$T_{\text{50\%}}$ and $T_{\text{75\%}}$. This means that the planet suddenly accreted enough mass to go from having less than $50\%$ 
of its final mass to have more than $75\%$ of its final mass. The same happens with planet a of the second set.}
\begin{center}
\begin{tabular}{|c|c|c|c|c|c|c|c|}
\hline 
\hline
Simulations  & Planet   &  $M$($M_\oplus$)  & $W$($\%$) & $t_{\text{50.class-D}}$ (Myr) & $t_{\text{50.class-W}}$ (Myr) & $T_{\text{50\%}}$ (Myr) & $T_{\text{75\%}}$ (Myr) \\
\hline
\hline
First Set   & \emph{b}  & 1.19             &  4.51  &    21.9      & 81.3               & 35                    & 35   \\
            & \emph{d}  & 0.66             &  8.10  &    21.8      & 67.3               & 2.9                   & 13.4 \\
\hline
Second Set  & \emph{a}  & 2.21             &  8.75  &    21.8      & 56.6               & 51.7                  & 51.7 \\  
            & \emph{c}  & 2.00             &  10.76 &    39.8      & 66.9               & 6                     & 20.5 \\ 
            & \emph{e}  & 1.37             &  7.49  &    27.1      & 67.8               & 4.4                   & 18.4 \\ 
\hline
\hline
\end{tabular}
\end{center}
\label{tab:3}
\end{table*}

%______________________________________________________________

\section{Discussion}

\begin{figure}[ht]
 \centering
 \includegraphics[angle=0, width= 0.4\textwidth]{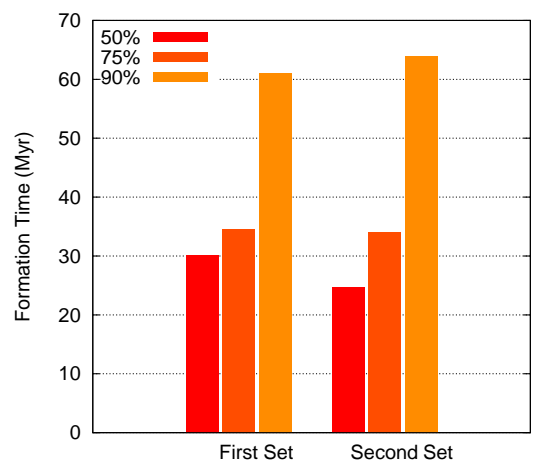}
 \caption{Timescales for planets in the HZ to reach a given fraction (50\%, 75\%, or 90\%) of their final masses for the set of
 simulations with ad hoc initial conditions and for the set of simulations with initial conditions from a semi-analytical model.
 These values are averaged for all the HZ planets in each simulation.}
 \label{fig:Tiempos}
\end{figure}

 We found that at the end of the evolution of our simulations, five and six planets are produced in the HZ of the system from the first and second set of initial 
conditions, respectively. For the first set, planets of the HZ present masses ranging from 0.66$M_{\oplus}$ to 2.27$M_{\oplus}$, with a median value of 
1.65$M_{\oplus}$. For the second set, the mass of the planets formed in the HZ ranges from 1.18$M_{\oplus}$ to 2.21$M_{\oplus}$, with a median value of
1.66$M_{\oplus}$. The timescales associated to the last giant impact received for the planets of the HZ are quite similar for both sets of simulations, 
with mean values of 35 Myr and 40 Myr for the first and second set, respectively. However, not always the last giant impact means that the planet 
presents the majority of it final mass. Figure \ref{fig:Tiempos} shows the mean time values that the planets in the HZ need to reach 50\%, 75\%, and 
90\% of their final mass for both sets. The values are averages for all planets in the HZ and, as we can see, the results are also quite similar for 
both sets.

A simple analysis of the feeding zones of the HZ planets (Figure \ref{fig:4}) reveals that the planets in the first set of simulations accrete a 
higher percentage of material from the outer disk that the potentially habitable planets produced in the second set. In fact, for the first set, 
the final water contents range from 4.5\% to 39.5\% by mass, with a median value of 32.5\%. For the second set, planets in the HZ have final water 
contents ranging from 7.5\% to 33.6\% by mass, with a median value of 17.5\%.

Our simulations form ``Earth-like planets'' and ``highly water-rich planets'' or ``water worlds'' in the HZ. All the Earth-like planets, in both sets, 
grow in situ and their feeding zones are restricted to the inner disk. 
Thus, their final water contents are not primordial, and a detailed analysis shows that planetesimals are the only population responsible 
for their final amounts of water. In addition, an analysis of the accretion timescales shows that all these planets accreted half of the class-W 
planetesimals after they have achieved $75\%$ of their final masses. Moreover, they continue accreting class-W planetesimals after 90 Myr of evolution. 
On the other side, the Earth-like planets in both sets of simulations accreted different ammounts of class-W planetesimals. Particularly, the mass 
of water accreted by the Earth-like planets of the second set is greater than the one accreted by the first set, by a factor of 1.9 - 4.  
The highly water-rich planets are planets whose accretion seeds come from the outer zone of the disk (planets a, c and e of the first set and planet 
d of the second set) or grow near their initial positions but have accreted embryos from beyond the snow line. The sourses of water suply are also 
different in both sets. Planetary embryos are the most responsible for the final water contents of the water worlds of the first set but for two of 
the three planets of the second set, the final water contents are provided by embryos and planetesimals in equal parts. 

However, it is worth mentioning that embryos
and planetesimals might not be the only source of water. \citet{Izidoro2013} suggested that a compound model that considers
water accretion through collisions of protoplanetary bodies and the contribution of the water locally adsorbed
from the nebula is more efficient in the amount and time of the delivery of water to Earth-like planets.

Considering more realistic initial conditions to include in the N-body simulations is important to get a more realistic accretion history of
the final planets. However, it is worth noting that the semi-analytical model used to generate these initial conditions contains several 
simplifications which might affect the results and could lead to different final configurations for the distribution of embryos and planetesimals.
Particularly, our model neglects the presence of embryo gaseous envelopes because the masses of the embryos are small enough not to 
produce major differences. Moreover, the semi-analytical model does not include the effects of planetesimal fragmentation or pebble accretion. 
On the other hand, we do not include type I migration in our model because many aspects of this phenomenon are still under debate. 
Computing the N-body interactions for the embryos during the gaseous phase could also change their spacial distribution. Regarding the planetesimal 
distribution, different sizes of planetesimals could lead to different final planetesimal surface densities.

A detailed knowledge about the accretion history of the planets of a given system is very important in order to analyze several
physical characteristics. In particular, all our simulations are developed assuming perfectly inelastic collisions that conserve mass
and water content. However, a more realistic treatment of the collisions can lead us to a better understanding about the differentiation of a 
planet, composition, formation of structures such as a core and a mantle, as well as to quantify its abundance of water \citep{Marcus2010, Chambers2013, Bonsor2015}.
At the same way, a better knowledge about the accretion history of a planet allows us to understand the evolution of its atmosphere
\citep{Schlichting2015}. In fact, impacts due to planetary embryos and planetesimals can generate a relevant atmospheric mass loss during
the formation and evolution process.   

\section{Conclusions}
We carried out N-body simulations aimed at studying the planetary formation process and water delivery
around Sun-like stars in low-mass disks and in absence of gas giants. The main goal of our investigation is to 
study the sensitivity of the results to the particular initial distribution addopted for planetesimals and planetary embryos 
after the gas dissipation. The first set of simulations was based on \citet{RoncodeElia2014} and assumes ad hoc
initial conditions while the second set of simulations derives initial conditions
from a semi-analytical model, which is capable of analyzing the evolution of a planetary system during the gas phase. 
Our main results suggest:
\begin{itemize}
\item The number of planets formed in the HZ of the system is not sensitive to the initial conditions proposed in both sets
of simulations. Moreover, the masses of the planets of the HZ do not show significant differences in both sets.

\item The main differences observed between both sets of simulations are associated to the accretion history of the planets of the HZ. Primarily, such discrepancies are related to the accretion of water-rich material.

\item Earth-like planets form in situ and their feeding zones are restricted to the inner region of the disk and these 
Earth-like planets present several Earth's water contents. 

\item Highly water-rich planets present different characteristics in both sets: these are planets whose accretion seeds come 
from the outer zone of the disk or are planets which grow near their initial positions but have accreted embryos from beyond 
the snow line, meaning that their feeding zones extend to the outer region of the disk.

\item Our simulations indicate that planetary embryos are the most responsible for the final water contents of the water 
worlds in the HZ of the first set, while planetesimals play a secondary role. Of the three water worlds formed in the HZ from 
the second set of initial conditions, only one of them shows a similar behavior to that previously described. However, the 
primordial water content of two of such planets is negligible and their final water contents are provided by impacts of 
embryos and planetesimals in equal parts.

\item Almost all the simulations formed Earth-like planets within the HZ. As they grow in situ, their final water contents are 
not primordial, and planetesimals are the only population responsible for their final amounts of water. In addition, all 
these planets accreted half of the class-W planetesimals after they have achieved $75\%$ of their final masses. Moreover, 
they continue accreting class-W planetesimals after 90 Myr of evolution. These results show that the water delivery is a late process 
in the evolution of an Earth-like planet and this is of great interest because it can facilitate 
the retention of water and other volatiles on the surface.
\end{itemize}

Finally we conclude that the inclusion of more realistic initial conditions in N-body simulations of terrestrial-type planet formation is crucial to do a more realistic analysis of the accretion history of the planets resulting from the formation of a planetary system.

%______________________________________________________________
\begin{acknowledgements}    
We thank the anonymous referee for constructive coments and we also thank the suggestions of the editor Tristian Guillot which helped us to improved 
the manuscript. This work was funded with grants from Consejo Nacional de Investigaciones Cient\'{\i}ficas y T\'ecnicas de la Rep\'ublica Argentina
and Universidad Nacional de La Plata (Argentina). 
\end{acknowledgements}

%-------------------------------------------------------------------

\bibliographystyle{aa} % style aa.bst
\bibliography{Paper2-version1} % your references Yourfile.bib

\end{document}